\documentclass[journal=nalefd,manuscript=letter,layout=traditional]{achemso}
\usepackage[version=3]{mhchem} % Formula subscripts using \ce{}
\usepackage[]{graphicx}
\usepackage{amsmath, amssymb}
\author{Sanghyun Jo}
\affiliation{DQMP and GAP, Universit\'{e} de Gen\`{e}ve, 24 quai
Ernest Ansermet, CH-1211 Geneva, Switzerland}
\author{Davide Costanzo}
\affiliation{DQMP and GAP, Universit\'{e} de Gen\`{e}ve, 24 quai
Ernest Ansermet, CH-1211 Geneva, Switzerland}
\author{Helmuth Berger}
\affiliation{Institut de Physique de la Matiere Complexe, Ecole Polytechnique Federale de Lausanne, CH-1015 Lausanne, Switzerland}
\author{Alberto F. Morpurgo}
\affiliation{DQMP and GAP, Universit\'{e} de Gen\`{e}ve, 24 quai
Ernest Ansermet, CH-1211 Geneva, Switzerland}
\email{Alberto.Morpurgo@unige.ch}

\title [] {Electrostatically induced superconductivity at the surface of WS$_2$}


\begin{document}


%\begin{tocentry}

%Some journals require a graphical entry for the Table of Contents.
%This should be laid out ``print ready'' so that the sizing of the

%Inside the \texttt{tocentry} environment, the font used is Helvetica
%8\,pt, as required by \emph{Journal of the American Chemical

%The surrounding frame is 9\,cm by 3.5\,cm, which is the maximum
%permitted for  \emph{Journal of the American Chemical Society}
%graphical table of content entries. The box will not resize if the
%content is too big: instead it will overflow the edge of the box.

%This box and the associated title will always be printed on a
%separate page at the end of the document.

%\end{tocentry}

\begin{abstract}
We investigate transport through ionic liquid gated field effect transistors
(FETs) based on  exfoliated crystals of semiconducting WS$_2$. Upon electron accumulation, at surface
densities close to --or just larger than-- 10$^{14}$ cm$^{-2}$,
transport exhibits metallic behavior,  with the surface resistivity decreasing
pronouncedly upon cooling. A detailed characterization as a function of temperature and
magnetic field clearly shows the occurrence of a gate-induced superconducting transition
below a critical temperature $T_c \approx 4$ K, a finding
that represents the first demonstration of superconductivity in tungsten-based semiconducting transition
metal dichalcogenides. We investigate the nature of superconductivity and find significant inhomogeneity,
originating from  the local detaching of the frozen ionic liquid from the WS$_2$ surface.
Despite the inhomogeneity, we find that in all cases where a fully developed zero resistance state
is observed, different properties of the devices exhibit a behavior characteristic of a
Berezinskii-Kosterlitz-Thouless transition, as it could be expected in view of the two-dimensional
nature of the electrostatically accumulated electron system.
\end{abstract}


Semiconducting transition metal dichalcogenides (TMDs; chemical
formula MX$_2$, with M = Mo, W and X = S, Se, Te) have been
recently attracting considerable interest. Being van der Waals
layered materials,  thin crystalline flakes can be extracted out
of bulk crystals by mechanical exfoliation, and allow the
realization of high-quality two-dimensional (2D) electronic
systems with unique
properties\cite{wang2012electronics,chhowalla2013chemistry}. From
a fundamental perspective, for instance, the absence of inversion
symmetry\cite{wang2012electronics,chhowalla2013chemistry,xiao2012coupled},
a strong spin-orbit
interaction\cite{wang2012electronics,chhowalla2013chemistry,xiao2012coupled},
and the direct nature of the
band-gap\cite{mak2010atomically,splendiani2010emerging} in
monolayers give unique control over the valley and the spin
degrees of
freedom\cite{xiao2012coupled,cao2012valley,zeng2012valley,
mak2012control,jones2013optical,mak2014valley,xu2014spin}.
Additionally, high-quality field-effect transistors
(FETs)\cite{radisavljevic2011single, zhang2012ambipolar,
braga2012quantitative, radisavljevic2013mobility,
baugher2013intrinsic, jo2014mono, ovchinnikov2014electrical,
gutierrez2014surface} and different kinds of opto-electronic
devices\cite{mak2014valley,jo2014mono,sundaram2013electroluminescence,
lopez2013ultrasensitive,
pospischil2014solar,baugher2014optoelectronic,ross2014electrically,ubrig2014scanning,
zhang2014electrically} have been demonstrated, which may be in
principle advantageous for future applications.

Another reason that has attracted attention to these materials is
the occurrence of  superconductivity upon accumulation of
electrons at the surface of MoS$_2$ and MoSe$_2$, as it has been
observed in a FET configuration\cite{ye2012superconducting,
taniguchi2012electric,shi2014MoSe2}. Key to this result is the use
of ionic liquid for electrostatic
gating\cite{panzer2005low,shimotani2006electrolyte}, which enables
density of carriers in excess of $n=10^{14}$ cm$^{-2}$ to be
accumulated at the material surface, a much larger value than that
accessible in FETs with conventional solid dielectrics. The
possibility to induce superconductivity electrostatically is
particularly appealing as compared to other methods (e.g., doping
by intercalation\cite{somoano1975alkaline,woollam1977physics})
because it does not cause structural modifications in the
material, and therefore it is likely to affect less its electronic
band structure (or at least to affect it in a more controlled way,
since the result of intercalation depends strongly on the spatial
arrangement of the intercalant atoms). Additionally, the technique
allows exploring superconductivity in carrier density ranges that
may not be accessible by intercalation, as it has been recently
reported for MoS$_2$ (where a lower carrier density unexpectedly
gives rise to a larger critical
temperature)\cite{ye2012superconducting}.

So far, only a very limited amount of work on electrostatically
induced superconductivity in ionic liquid gated TMDs has been
reported\cite{taniguchi2012electric,ye2012superconducting,shi2014MoSe2},
and many different interesting questions remain to be answered.
One is whether the possibility to induce superconductivity
electrostatically is confined to Mo-based compounds (the only ones
in which the phenomenon has been observed so far), or whether it
is more general. Indeed, recent low-temperature transport
measurements of ionic liquid gated WSe$_2$ FETs have not revealed
the occurrence of a superconducting transition for $T$ as low as
1.5 K, despite the accumulation of surface electron densities
significantly larger than $\sim$ 10$^{14}$
cm$^{-2}$\cite{yuan2013zeeman}. Other questions concern the nature
of superconductivity, specifically the homogeneity of the
superconducting state, and whether signatures can be found in the
experiments of the 2D character\cite{berezinskii1972destruction,
kosterlitz1973ordering, kosterlitz1974critical,
beasley1979possibility,doniach1979topological,
halperin1979resistive,benfatto2009broadening} of the
superconducting transition, which can be expected given that at
$n\sim 10 ^{14}$ the thickness of the electron accumulation layer
at surface\cite{horowitz2000temperature} is approximately 1 nm.

We have addressed these issues by performing investigations of
low-temperature transport in ionic liquid gated FETs realized on
thin exfoliated crystals of WS$_2$. At sufficiently high gate
voltage --corresponding to accumulated surface carrier density of
$n\simeq 10^{14}$-- we observe that the square resistance
initially decreases upon cooling, and then saturates when $T$ is
lowered below 40 K, a clear manifestation of metallic behavior.
Upon further cooling, a sharp decrease in resistivity is observed
with an onset at approximately 4 K, leading to a zero resistance
state at $T\simeq 0.5 K$ (the precise values of temperature depend
on the device and on the cool-down). The application of a
perpendicular magnetic field $B$ results in the destruction of the
zero-resistance state (already at very low $B$) and in a
progressive increase of the device resistance, which at $B\simeq
0.1$ T reaches the normal state value. These observations provide
a clear experimental demonstration of the occurrence of
electrostatically induced superconductivity at the surface of
WS$_2$.

We further investigate the properties and character of the
superconducting state, by performing  measurements on devices with
many different pair of contacts, which allow us to characterize
transport locally. Although, a clear signature of the
superconducting transition is generally observed irrespective of
the part of the device probed, a strong inhomogeneity in square
resistance, carrier density and critical temperature are also
regularly observed. We attribute the inhomogeneity to the frozen
ionic liquid that locally detaches from the WS$_2$ surface,
resulting in large variation of the local capacitance, and
therefore of the density of accumulated charge. Despite these
inhomogeneities, we find that the temperature-dependent
$I-V$ curves measured in the regions of the devices where a zero-resistance state
is observed exhibit the behavior typically associated to 2D
superconductors\cite{berezinskii1972destruction,
kosterlitz1973ordering, kosterlitz1974critical,
beasley1979possibility,doniach1979topological,
halperin1979resistive,benfatto2009broadening} of the
Berezinskii-Kosterlitz-Thouless (BKT) type. Besides providing the
first demonstration of electrostatically induced superconductivity
at the surface of WS$_2$, our findings reveal different
interesting aspects of ionic-liquid gated devices that are
certainly relevant to understand the low-$T$ normal and
superconducting transport properties of these systems.

Fig. 1(a) shows an optical microscope image of a 20 nm thick
WS$_2$ flake. The flake was exfoliated from a bulk crystal and
transferred onto a Si/SiO$_2$ substrate. Contacts consist of a
Ti/Au (10/50 nm) bilayer and were defined by conventional
nano-fabrication techniques (electron-beam lithography, metal
evaporation and lift-off). Together with the contacts, a
large-area pad acting as gate was also defined (see Fig. 1(b) for
the schematic device configuration). To form the ionic gated FET,
a small droplet of ionic liquid DEME-TFSI was placed onto the
device (Fig. 1(b)) in the controlled environment of a glove-box,
after which the device was rapidly transferred in the vacuum
chamber of a 3He cryostat used for the $T$-dependent transport
measurements. Before starting the measurements, the device was
left in vacuum, at room temperature, at a pressure of $\approx
10^{-6}$ mbar for one day to remove humidity and oxygen present in
the ionic liquid. Three similar devices were investigated in
detail, with multiple cool-downs at different values of gate voltage,
and exhibited virtually identical transport behavior; here
we show data from one of them which is representative of this
behavior.

We start with the characterization of the device room-temperature
transport properties. The transfer curve (i.e., the source-drain
(S-D) current $I_{SD}$ measured as a function of gate voltage
$V_G$) is shown in Fig. 1(c) for a source-drain bias $V_{SD}$ = 1
mV. The data exhibit clear ambipolar transport, similarly to what
we have reported previously\cite{braga2012quantitative}. To
eliminate contact effects and to check the device homogeneity, we
measured  four-terminal resistances using as voltage probes
different, adjacent pairs of contacts (numbered from 1 to 10 in
Fig. 1(a); in all cases --here and in the remainder of this paper,
if not stated otherwise-- the current is sent from the S to the D
contact). The square resistance values ($R^{\square}$) obtained in
this way are plotted in the inset of Fig. 1(c) as a function of
$V_G$. They exhibit variations of at most 50 $\%$, in large part
due to the difficulty to define the device dimensions precisely
when calculating $R^{\square}$, and to the geometry of the WS$_2$ flake,
that results in a non perfectly uniform current distribution. Overall,
therefore, these measurements indicate a reasonably good
electrical homogeneity of the devices at room temperature, when
the ionic gate is still in the liquid state.

To investigate low-temperature electronic transport upon
accumulation of a high density of electrons, the device
temperature was set just above the freezing point of
DEME-TFSI\cite{zhang2012ambipolar,jo2014mono,sato2004electrochemical},
and the gate voltage applied. At this lower $T$ possible chemical
reactions between WS$_2$ and the ionic liquid slow down, enabling
a much wider range of $V_G$ to be applied without causing device
degradation. Values of $V_G$ between 3 and 6 V were applied in
different measurement runs, after which the device was cooled down
to the base temperature of our 3He system (0.25 K). Throughout
this procedure, the gate leakage current was monitored to ensure
that it remained negligibly small (below 1 nA).

Figure 2(a) shows the temperature dependence of the square
resistance at $V_G =$ 3.7 V, obtained in a four-probe
configuration by measuring the voltage between contacts 7 and 8
(see Fig. 1(a); a current bias of 30 nA was forced from S to D).
Metallic transport is observed, with $R^{\square}$ first
decreasing as $T$ is lowered to approximately 30-40 K, and then
saturating to $R^{\square}\simeq 550 \Omega$. Upon further cooling
we see that, below approximately $T=4$ K, the device exhibits a
sharp, large decrease of resistance. Fig. 2(b) zooms-in on the
behavior of $R^{\square}(T)$ below $T =$ 10 K, making it apparent
that at $T \simeq 4$ K, the device starts undergoing a transition
to a zero resistance state, which is attained below $T =$ 0.4 K.
This is the first observation of a superconducting state in
WS$_2$. It demonstrates that the possibility of electrostatically
inducing superconductivity in semiconducting TMDs is not confined
to Mo-based
compounds\cite{taniguchi2012electric,ye2012superconducting,shi2014MoSe2}
(the only ones in which the phenomenon had been reported so far).

To further substantiate that the zero-resistance state is a
manifestation of superconductivity, we investigated transport in
the presence of an applied perpendicular magnetic field. The
magnetoresistance measured at $T =$ 0.25 K is shown in the inset
of Fig. 2(b). We find that a truly zero-resistance state persists
only up to $B =$ 10 G, and that at higher values of $B$ the
resistance increases rapidly, reaching the value measured in the
normal state at $B =$ 0.14 T (i.e., the critical field at $T=0.25$
K is $B_{C}=0.14$ T). For large field values the magnetoresistance
is essentially negligible. Fig. 2(c) further shows the evolution
of $R^{\square}$ as a function of $T$, in the presence of
different values of applied perpendicular field, which confirms
how superconductivity is  suppressed on  a magnetic field
scale of $B \approx 0.1$ T. The same data show that above 4 K
virtually no change of resistance with magnetic field is observed,
confirming that the critical temperature $T_C \approx$ 4 K (rather
surprisingly, $T_C$ does not seem to shift significantly upon the
application of a magnetic field; this may be related to the
inhomogeneity present in the device, see below). To identify $T_C$
more precisely, we subtract the resistances measured at $B =$ 0
and 0.1 T (inset of Fig. 2(c)), from which we can see that the
difference of the two resistances starts to deviate from zero for
$T< 4.25$ K (i.e.,  $T_C=4.25$ K).

We have also analyzed the current-voltage ($I-V$) characteristics,
obtained by measuring voltage between contacts 7 and 8 as a
function of $I_{SD}$. Fig. 2(d) and 2(e) show the evolution of the
$I-V$ curves as a function of temperature (at $B =$ 0 T; Fig.
2(d)) and of magnetic field (at $T =$ 0.25 K; Fig. 2(e)). At $T =$
0.25 K with $B =$ 0 T, the maximum supercurrent (i.e., the
critical current) is $I_{SD} =$ 0.3 $\mu$A. With increasing $T$ or
$B$ the supercurrent is suppressed, and the $I-V$ curves become
fully linear for $T>T_C$ or $B>B_C$, as expected. In Fig. 2(f),
the evolution of critical current with magnetic field is shown
more clearly in a two-dimensional (2D) color plot of the $I$-$|V|$
characteristics measured at base temperature as a function of $B$.

Having established the occurrence of an electrostatically induced
superconducting state in ionic liquid gated WS$_2$ devices, we
discuss some important aspects of its nature that could be
determined from our experiments. A first, very important aspect
regards the presence of a rather strong electronic inhomogeneity,
which can be illustrated by several different measurements. For
instance, Fig. 3(a) shows the low-temperature square resistance
$R^{\square}(T)$ at $V_G =$ 3.7 V obtained by using contacts S and
D to current bias the device ($I_{SD} =$ 30 nA), and different
pairs of contacts (indicated according to the labeling of Fig.
1(a)) to measure the resulting voltage drop. In all measurements
the superconducting transition is clearly seen. However, the
square resistance in the normal state, the value of critical
temperature, and the resistance well below the superconducting
transition all depend strongly on the contacts used to perform the
measurements. The square resistance varies by almost one order of
magnitude (in contrast with the rather homogeneous behavior
observed at room temperature), $T_C$ ranges between approximately
2 and 4 K (the range is even broader if we include data from all
the device measured), and the remnant square resistance well below
$T_C$ varies between 0 to approximately 300 $\Omega$. This type of
behavior has been observed in all the different devices that we
have measured, and upon cooling a same device multiple times. The
inhomogeneity also manifests itself in large local variations of
carrier density (approximately a factor of 4 across the entire
device) extracted from Hall measurements performed with different
pairs of contacts (see Fig. 3(b)). From all these observations we
conclude that the electronic inhomogeneity needs to be taken into
account in the interpretation of the experimental results. For
instance establishing a well-defined relation between $T_C$ and
$n$ is problematic, as shown in Fig. 3(c): if we plot all the data
points available, any possible relation between $T_C$ and $n$ is
eclipsed by the device inhomogeneity (establishing such a relation
may be possible by selecting data according to some criterion, but
it certainly remains important to check whether the result
represents well the behavior of the experimental system).

In the superconducting state, the inhomogeneity also manifests
itself in interesting ways that provide information about the
underlying length scale. Fig. 3(d) shows the variation of $I-V$
curves upon finely varying the applied perpendicular magnetic
field, on a scale of a few Gauss. An oscillatory behavior of the
critical current is apparent, with a rather precisely well-defined
period. These oscillations in critical current are a manifestation
of quantum interference, the same mechanism responsible for the
operation of SQUID devices\cite{tinkam2004introduction}. Their
observation implies the presence of a superconducting loop
encircling a non-superconducting region, with the critical current
responding periodically (with a period given by the
superconducting flux quantum $h/2e$) to variations of the magnetic
flux through the loop. From the period measured experimentally
(0.36 Gauss) we estimate the area of the loop to correspond to a
square of 7.6 $\times$ 7.6 $\mu$m$^2$ (compatible with the size of
the device area probed by the four-terminal measurements, which is
approximately 25 $\times$ 15 $\mu$m$^2$). This observation --as
well as all the measurements discussed above-- are compatible with
the inhomogeneity occurring over a rather large length scale
(several microns; it appears possible that on a length scale of
approximately 1 micron the homogeneity is significantly better).
In the future it will be particularly important to determine the
length scale associate to the inhomogeneous charge density in
ionic liquid gated devices and find ways to substantially suppress
this inhomogeneity.

We believe that the observed inhomogeneity originates from the
frozen ionic liquid detaching locally from the surface of the
WS$_2$ device (e.g., because of different thermal expansion
coefficients), with the resulting increased distance between
frozen liquid and the WS$_2$ surface causing a large decrease of
geometrical capacitance, and correspondingly, of induced charge.
Many observations support this idea. For instance, we found that
measurements of all the four-terminal resistances in a device such
as the one shown in Fig. 1(a) allow us to determine (at least
approximately) the path followed by the current flowing from the S
to the D contact, and provide clear evidence for large patches of
the surface not being conducting. Another indication can be
obtained by comparing the temperature dependence of the square
resistance upon cooling and warming up the device (see Fig. 3(e)).
Upon warming up the device, the resistance is significantly larger
than that measured during cooling down (as a result of the
detachment of the liquid), but as soon as the melting temperature
of the liquid is reached (at approximately 200 K), the  resistance
recovers its original value. Indeed, as we discussed earlier on,
when the ionic liquid is kept above its freezing point, the
transport properties are rather homogeneous.

Another question worth investigating is whether the experiments
provide evidence for the 2D character of the superconducting
state, which can certainly be expected, since the thickness of the
accumulation layer at carrier density values of $n \approx
10^{14}$ cm$^{-2}$ --determined by the electrostatic screening
length-- is only approximately 1 nm. To address this issue, we
compare the behavior of the superconducting transition to
theoretical expectations for 2D Berezinski-Kosterlitz-Thouless
(BKT) superconductivity\cite{berezinskii1972destruction,
kosterlitz1973ordering, kosterlitz1974critical,
beasley1979possibility,doniach1979topological,
halperin1979resistive,benfatto2009broadening}. One of the
well-known properties of BKT superconductors is the occurrence of
power law $I-V$
curves\cite{halperin1979resistive,benfatto2009broadening}: $V
\propto I^\alpha$ where $\alpha =$ 1 for $T$ well above $T_{BKT}$,
$\alpha =$ 3 for $T = T_{BKT}$, and $\alpha >$ 3 for $T < T_{BKT}$
($T_{BKT}$ is the BKT transition temperature below which a finite
stiffness of the phase of the superconducting parameter appears;
it is smaller than the mean-field critical
temperature\cite{tinkam2004introduction} where superconductivity
appears locally, without global phase coherence). Fig. 4(a) shows
the $I-V$ curves measured well below the onset of
superconductivity in log-log scale: the linear behavior observed
(over one-to-two decades in current) reveals that a power law
dependence does indeed hold. From the slope we extract the
exponent $\alpha$, whose $T$-dependence is shown in Fig. 4(b).
$\alpha=3$ at $T=0.63$ K, providing a first estimate of $T_{BKT}$.
More evidence for BKT behavior is revealed by the temperature
dependence of the resistance close to $T_{BKT}$. Theoretically, it
is expected that $R = R_0 exp(-bt^{-1/2})$, with $R_0$ and $b$
material dependent parameters, and $t =T/T_{BKT} -
1$\cite{halperin1979resistive,benfatto2009broadening}. By plotting
$[d(\ln R^\square)/dT]^{-2/3}$ and extrapolating linearly this
quantity to zero provides a second, independent way to determine
$T_{BKT}$ (see Fig. 4(c)). Following this procedure we obtain
$T_{BKT} =$ 0.67 K, in fairly good agreement with our previous
estimate, indicating that at this level of analysis the
experimental data are compatible with superconductivity having a
2D character.

Finally, we briefly compare electrostatically induced
superconductivity on WS$_2$ to the same phenomenon observed on
MoS$_2$, as reported in Ref. \citenum{ye2012superconducting}.
Overall, superconductivity in MoS$_2$ appears to be much more
robust, with a significantly higher critical temperature (in
excess of 10 K), surviving up to much larger magnetic field values
(approximately 10 T at low temperature), showing critical currents
orders of magnitude larger (Costanzo \emph{et al.}, unpublished),
and exhibiting a considerably narrower width of the
superconducting transition as a function of temperature. Indeed,
we have observed superconductivity in MoS$_2$ also in our
laboratory, exhibiting features similar to those reported in Ref.
\citenum{ye2012superconducting}. Similar to the findings discussed
here, however, our measurements on MoS$_2$ also indicate the
presence of inhomogeneity on a macroscopic scale, and a behavior
compatible with 2D BKT superconductivity (our measurements on
MoS$_2$ will be presented elsewhere).

In summary, we have investigated electronic properties of ionic
liquid gated WS$_2$ exfoliated crystals and observed the first
occurrence of superconductivity in this material. Next to
characterizing the superconducting properties, we have found that
ionic liquid gated devices at low temperature exhibit a strong
inhomogeneity that needs to be taken into account when analyzing
the data, and observed that the measurements appears to be
compatible with the behavior expected for 2D BKT superconductors.

\begin{acknowledgement}
We gratefully acknowledge T. Giamarchi and Y. Iwasa for
discussions, and A. Ferreira for technical assistance. Financial
support from the Swiss National Science Foundation, and from the
EU Graphene Flagship is also gratefully acknowledged.
\end{acknowledgement}

%\begin{suppinfo}

%This will usually read something like: ``Experimental procedures and
%characterization data for all new compounds. The class will
%automatically add a sentence pointing to the information on-line:

%\end{suppinfo}


\bibliography{WS2_Superconductivity_Bib}

\begin{figure}
  \includegraphics[width=.9\textwidth]{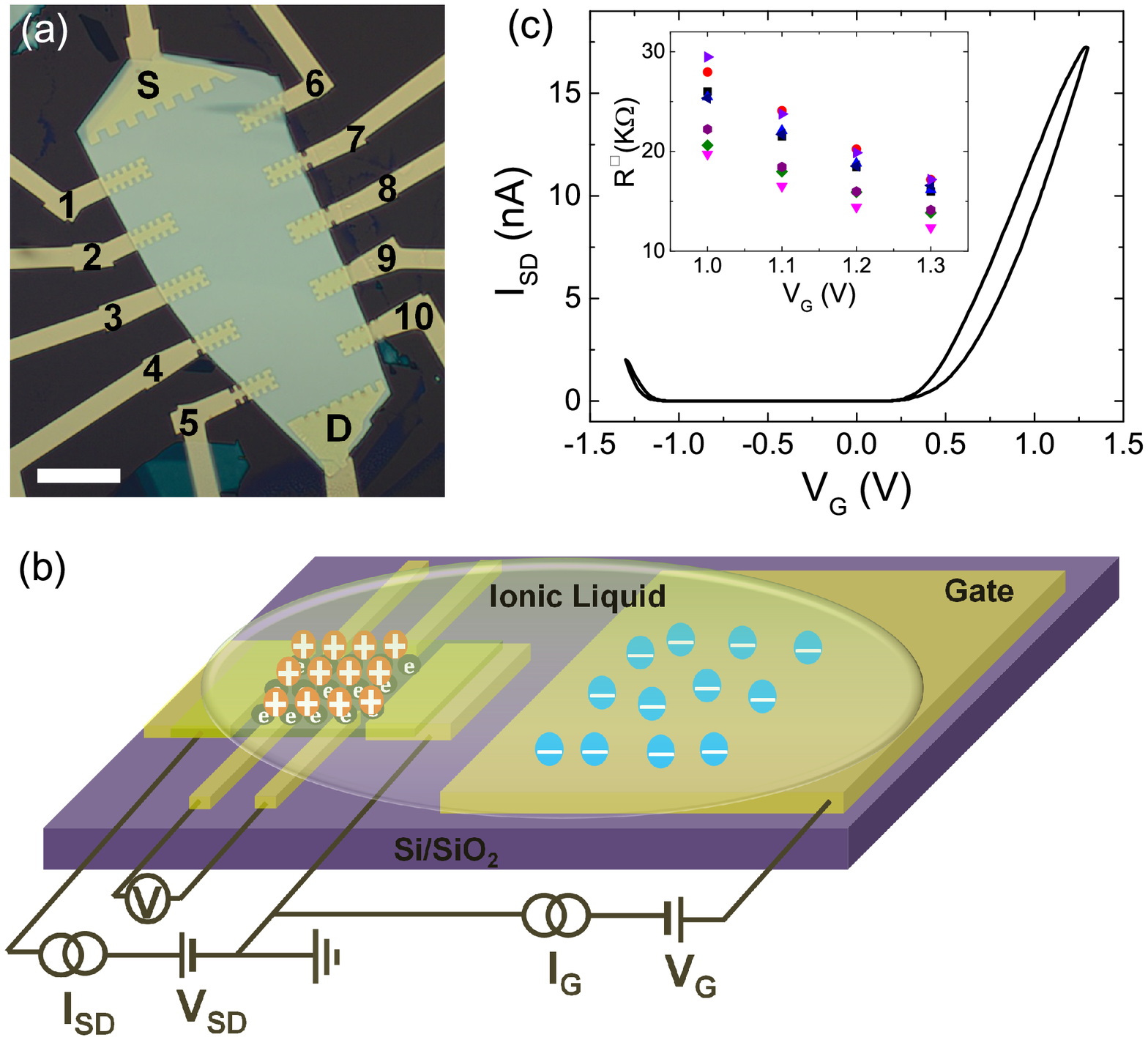}
  \caption{(a) Optical microscope image of a device used in our investigation (the WS$_2$ flake is 20 nm thick; the white bar is 15 $\mu$m long).
  The numbers (1 to 10) label the contacts used to probe the transport locally in the device, upon sending current from the source (S) to the
  drain (D) contact. The saw-tooth shape of the contacts is designed to increase the contact area, which helps to reduce the contact resistance. (b) Schematic illustration of our ionic liquid gated field effect transistors, under electron accumulation.
 (c) Source-drain current ($I_{SD}$) as a function of gate voltage ($V_G$) measured at room temperature with $V_{SD} =$ 1
 mV.
  The inset shows the square resistance extracted from multi-terminal measurements performed using different contacts as voltage probe
  (1 and 2, 2 and 3, \ldots, 9 and 10).}
  \label{fgr1}
\end{figure}

\begin{figure}
  \includegraphics[width=.9\textwidth]{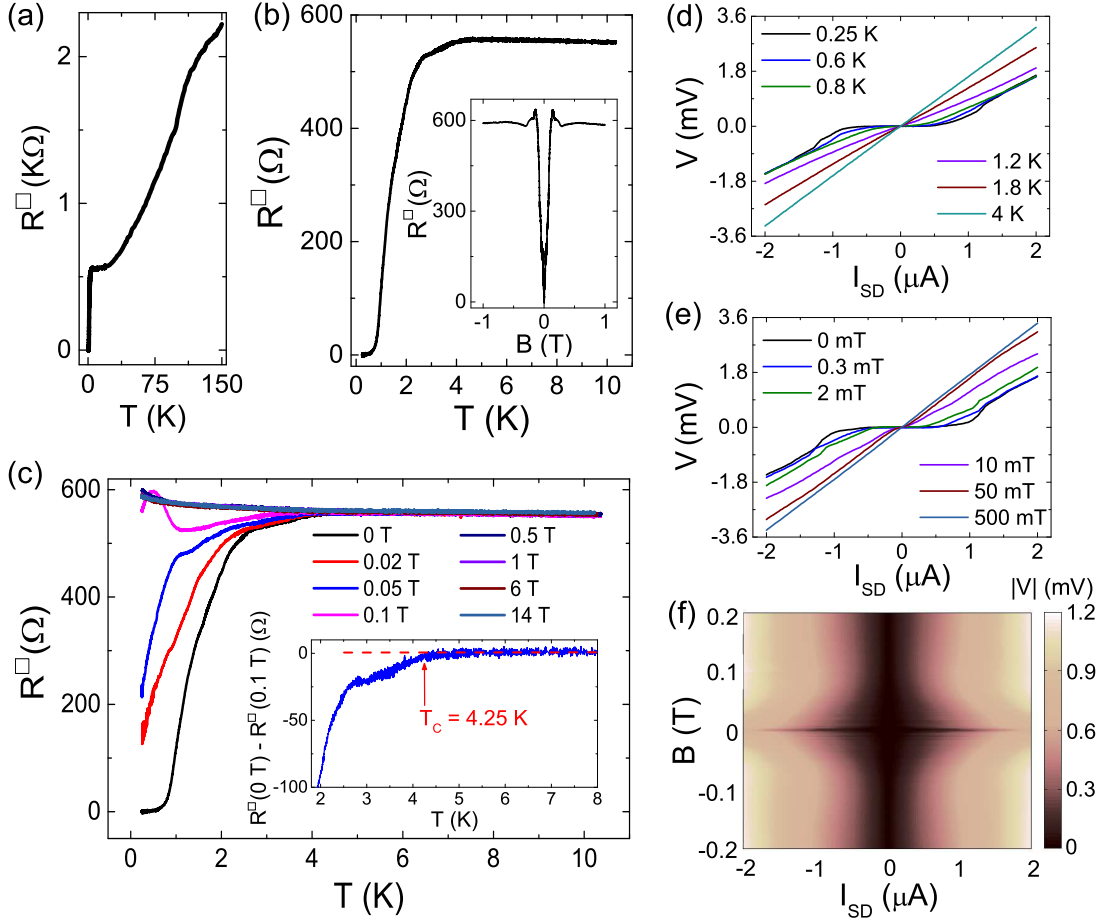}
  \caption{Four-probe transport properties recorded at $V_G$ = 3.7 V by measuring the voltage between contacts 7 and 8, as a function
  of source-drain current $I_{SD}$. The observed temperature dependence of the square resistance reveals metallic behavior at higher temperatures (a), followed by
  a superconducting transition at lower $T$ (b). The inset of panel (b) shows the magnetoresistance measured at $T =$0.25 K. (c) Evolution of the $T$ dependence of the
  square resistance, for different values of perpendicular magnetic field. The inset shows the quantity $R^\square$($B =$ 0 T) $-$ $R^\square$($B =$ 0.1 T), which
  makes it possible to determine $T_C$ more precisely. Evolution of current-voltage ($I-V$) characteristic as a function of $T$ at $B=0$ T (d) and as a function of $B$
  for $T =$0.25 K (e). (f) Two-dimensional color plot of $I-|V|$ as a function of $B$.}
  \label{fgr2}
\end{figure}

\begin{figure}
  \includegraphics[width=.9\textwidth]{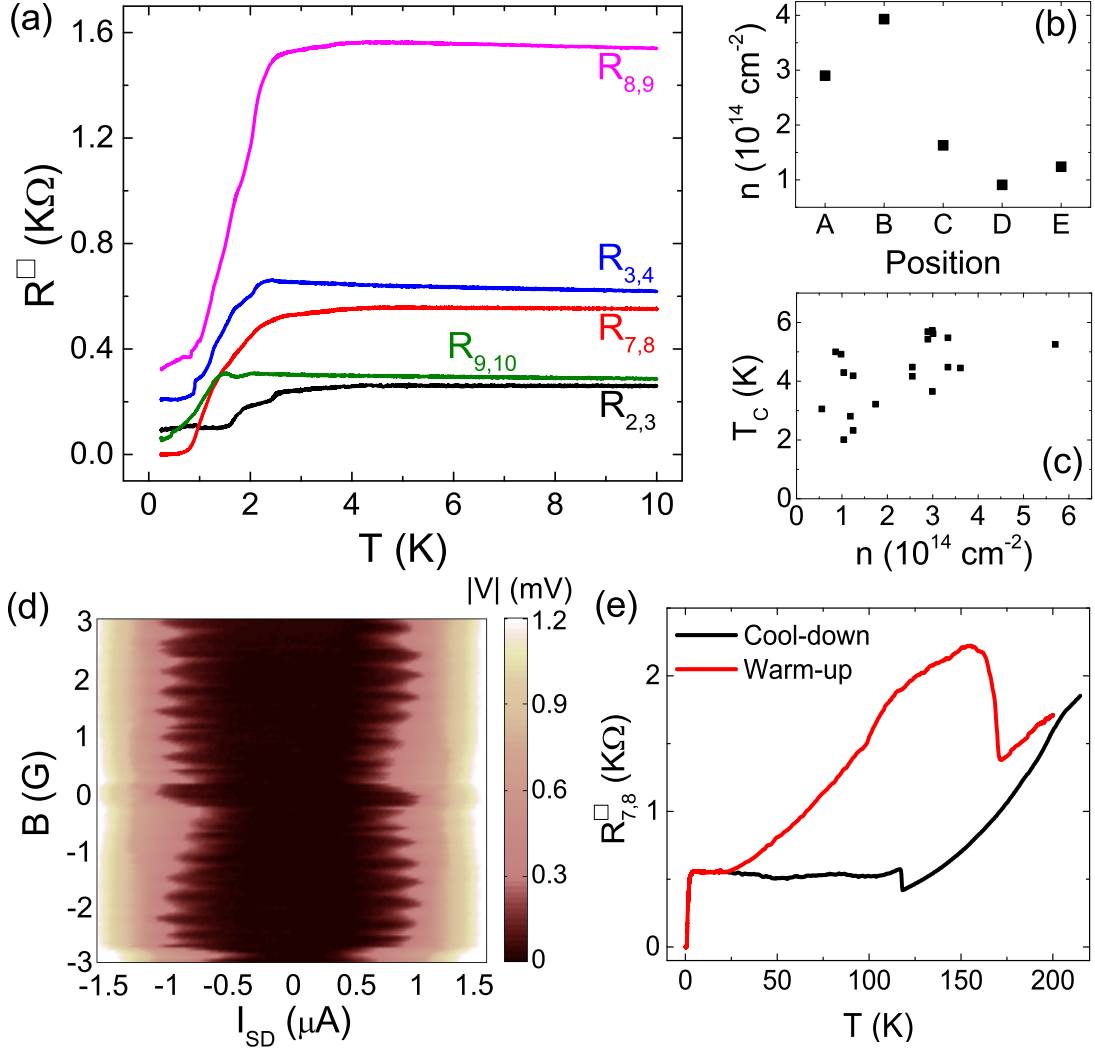}
  \caption{(a) Temperature dependence of the square resistance ($R^\square(T)$) measured in a four-terminal configuration, with different pairs
   of voltage probes ($V_G =$ 3.7 V; $I_{SD} =$ 30 nA), showing a considerable inhomogeneity. (b) Carrier density $n$ extracted at $V_G =$ 3.7 V from the measurement
   of the Hall effect at different positions in the device (each letter corresponds to measurements done with a different pair of contacts: A= contacts 1 and 6; B= 2 and 7; C= 3 and 8, D= 4 and 9, and E= 5 and 10, with the labeling of Fig. 1(a)). (c) $T_C$ as a function of $n$, showing data obtained with different electrode pairs in the three different devices that we have studied. The large variations of $n$ eclipses any systematic relation. (d) Color plot of the $I-|V|$ characteristic measured
   at low magnetic field $B$, showing a clear periodic oscillation of the critical current. (e) Square resistance $R^\square(T)$ measured as a function of $T$ while cooling and warming up the device.}
  \label{fgr3}
\end{figure}

\begin{figure}
  \includegraphics[width=.9\textwidth]{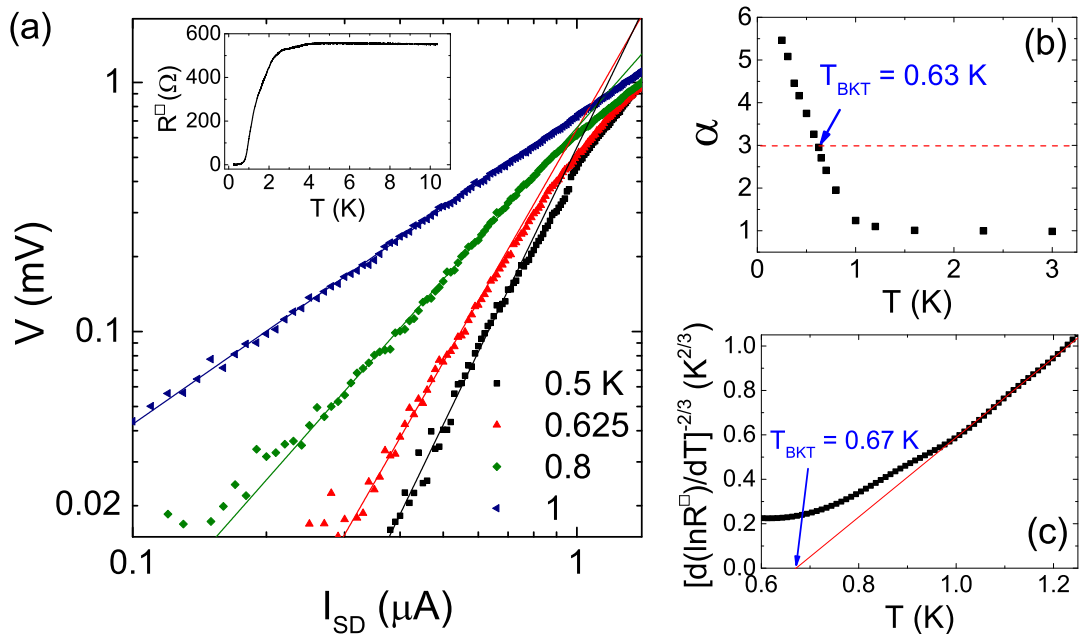}
  \caption{(a) Power law dependence $V \propto I^\alpha$ of the $I-V$ characteristics ($V$ measured with the contacts of 7 and 8 as a function of  $I_{SD}$) for different values of $T<T_C$. The solid lines are best fits done to extract the exponent $\alpha$ (the inset shows  $R^\square(T)$ measured with the same contacts). (b) Temperature dependence of the exponent $\alpha$. $T_{BKT}=0.63$ K is determined by the condition $\alpha=3$. (c) Plot of [d(ln$R^\square$)/d$T$]$^{-2/3}$, with a linear extrapolation of the linear part, which provides an independent determination of $T_{BKT}$. This method gives $T_{BKT} =$ 0.67 K, in fairly good agreement with the analysis of the $I-V$ curves.}
  \label{fgr4}
\end{figure}

\end{document}